\documentclass[twocolumn,twoside,slac]{revtex4}
\usepackage{graphicx}
\usepackage{fancyhdr}
\begin{document}

\title{BlueOx: A Java Framework for Distributed Data Analysis}

%

\author{J. Mans and D. Bengali}
\affiliation{Princeton University, Princeton, NJ 08544, USA}

\begin{abstract}
High energy physics experiments including those at the Tevatron and the 
upcoming LHC require analysis of large data sets which are best handled by
distributed computation.  We present the design and development of a 
distributed data analysis framework based on Java.  Analysis
jobs run through three phases: discovery of data sets available, 
brokering/assignment of data sets to analysis servers, and job execution.  Each
phase is represented by a set of abstract interfaces.  These interfaces allow
different techniques to be used without modification to the framework.  For
example, the communications interface has been implemented by both a packet 
protocol and a SOAP-based scheme.  User authentication can be provided either
through simple passwords or through a GSI certificates system.  
Data from CMS HCAL Testbeams, the L3 LEP experiment, and a hypothetical high-energy
linear collider experiment have been interfaced with the framework.  
\end{abstract}

\maketitle

\thispagestyle{fancy}

\section{Introduction}

Technology improvements over the last twenty years have allowed scientists to acquire and retain an 
ever-increasing amount of data from their experiments. This trend has the potential to greatly
increase the reliability of experimental results, to allow the detection of physical effects
which are extremely weak or rare, and to provide a library of data for testing new results and
theories. An increase in the amount of data can provide a lower statistical uncertainty in
results and allow a better evaluation of potential systematic effects.  Additionally, an increase in stored
data allows scientists to search for physical effects which may occur very rarely, since the
chance of seeing the effect is better with many observations. Finally, scientists do not have
to save only that data which is most interesting now -- data which might be interesting in the
future can also be stored.

The rapid decline in data storage costs has prompted many experimental groups to store vastly larger
quantities of data than before. For some types of experiment, the ability to analyze data has not kept
pace with the ability to record data. Consider the problem facing particle physicists at the
new Large Hadron Collider (LHC): Studies of the CMS experiment predict that the experiment
will generate 1 petabyte of data per year\cite{cms}. 
This is an amount of
data which large laboratories such as CERN and Fermilab struggle to manage and store, and
which individual universities have no hope of storing locally. However, the majority of researchers
in these experiments will wish to remain at their local universities where they can fulfill 
their teaching responsibilities and involve
undergraduates in the research process. This challenge is already confronting researchers in
the CDF and D\O\ experiments at Fermilab, where the Tevatron is currently generating large
volumes of data with even higher data rates to come.

An alternative to bringing the data to the analysis program at the local university is to
carry the analysis program to the data. Previously, this was usually done by having several
powerful machines at the experiment site which all researchers could use. A researcher would
compile and run his or her analysis on one of these central machines and the access to the
data would occur at high rates from the central store. This technique works well, but it is
hard to scale up to the hundreds or thousands of physicists in today's largest collaborations.
It also ignores the substantial computing resources which are available at other national and
university sites around the world. Most experimental collaborations keep archives of the data
at a central location but also distribute the data to many worldwide locations, from national
laboratories to regional centers to university group machines.

Such a data distribution carries many advantages. One advantage is increased system
robustness. For example, all work does not have to stop for remote users if there is trouble
with the network link between the United States and Geneva, Switzerland, where CERN is
located. Physicists can simply access the data from another source. Another advantage is the
speed of response and scalability. For most requests, the data should be available locally or
regionally, and the central servers will only be burdened by special requests. Despite its
advantages, a wide distribution of data breaks the traditional technique of bringing the
program to the data. Now, for many analyses the relevant data is not ``local'' anywhere, but
rather is spread over many different locations. The challenge is figuring out how to break up
the program so that it can follow the data in its distribution around the world, while
providing debugging and monitoring tools which are easy to use.

We have addressed this challenge by designing an analysis framework which aims to allow a
physicist to design a nearly-arbitrary analysis program and have the program executed
efficiently and transparently on a large distributed set of data. We based our design on the
needs of a physicist who wants to analyze the data. Usually, such a physicist is attempting to
achieve a pure sample of some interesting process and either study a feature of the process or
simply demonstrate the process's existence. The analysis contains a set of criteria
which select preferentially the events which match the signature of the desired process,
generally designated as the signal process, and remove all the other events from data, which
are designated as background. The separation techniques vary from the simple to the complex.
Sometimes the separation is done by setting upper and/or lower limits on quantities in the
events. Other analyses might train or use a neural network, build a multi-dimensional
probability distribution, or even implement an entirely new analysis technique which was not
conceived of when the framework was designed.  We have designed our framework to support both kinds
of analysis as completely as possible.

We have dubbed our project ``BlueOx'' after Paul Bunyan's companion, Babe the blue ox. BlueOx is
based on a set of autonomous agents which represent the physics users and which interact with
a set of servers to provide execution for physics analysis jobs. Our system is a framework,
which means that a physics analysis code is not a complete program but rather plugs into the
framework. The framework is responsible for ensuring that the analysis code is presented
with the right data at the right time and that the results are all gathered together
correctly. The role of the BlueOx framework is to provide a transparent service where an
analysis job which requires access to many data sets is broken into many sub-jobs and each
runs on a separate server.

The BlueOx framework is written and based on the Java$^{\scriptsize {\textrm{TM}}}$ programming
language. We selected Java specifically for two major advantages in distributed computing.
First, the Java compiler produces machine code which is not specific to any processor or
operating system, but rather runs within a ``virtual machine'' (VM)\cite{java}. The VM can be easily
emulated on any real computer, and compiled ``binaries'' will run without being recompiled.
Native code and libraries can be used, through the Java-Native Interface (JNI), but such usage
limits the portability of the code. Second, the Java VM provides a ``sandbox'' for the analysis
code. The server classes run outside the sandbox, so that they have full access to the system.
Inside the sandbox, analysis code and any classes downloaded from the user's computer will not
be able to read or write files, open network connections, or carry out any other dangerous
behaviors. The sandbox will protect the server against both malicious attacks and poorly
written analysis code.

\section{Design of the BlueOx Framework}

\subsection{Flow of a Job}

The execution of a job proceeds through three distinct phases: discovery, brokering, and 
execution. Each of these phases is represented by a separate set of abstract interfaces in 
BlueOx. 

1. {\bf Discovery } -- In the discovery phase, the agent obtains a list of data sets available on
any of 
the servers on the network. This list may be a complete list of all data sets or it may be 
based on a query submitted by the user. The list of data sets is provided without reference 
to which server or servers may actually host the data sets. The Discovery interface also 
allows access to data set descriptor objects, which provide a textual description of the data 
set, a count of events contained in the data set, and a map of additional parameters defined 
by the data set. The agent may be configured to use any number of different Discovery 
implementations. The implementations written to date are discussed below.

2. {\bf Brokering } -- In the brokering phase, each data set the user wishes to consider in the 
analysis is assigned to a server. 
The brokering phase begins when the user gives the list of chosen data
sets and the analysis code to the agent. As with Discovery, the agent may
be configured to use any given implementation of the Brokerage interface.
This interface accepts a list of data sets or data set 
descriptors and returns a list of contracts. Each contract specifies a server and a list of 
data sets which are assigned to that server for analysis. 

3. {\bf Job execution} -- Using the contracts obtained from the brokering phase, the agent splits 
the job and distributes it to the servers. The servers can download any required analysis 
classes from the agent. The agent is responsible for monitoring the job execution and 
attempting to re-broker parts of jobs which fail due to a server or network malfunction. Once 
the job is complete, the agent retrieves the split copies of the job and merges them together 
before returning the results to the user.

\subsection{Abstract Interfaces}

BlueOx operates through sets of abstract interfaces.  For each phase of BlueOx operation, one
or more abstract interfaces comes into operation, allowing many different data stores, communication 
schemes, and user authentication techniques to be used with the framework.  

The data handling interfaces allow the BlueOx server to open and access events from data stores.  Each
data store may contain any number of data sets, which are identified by unique names.  The DataSet
abstract interface allows the server to iterate over the events in a data set or to access individual
events as identified by sixty-four bit numbers which are required to be unique within the data set.
Each data set must be able to produce a DataSetDescriptor object which contains the unique name of the 
data set, the number of events, and a list of descriptive parameters which can contain anything additional the data set
wants to provide.
In all, the data handling interfaces define thirteen methods sufficient to integrate a data source into 
the BlueOx framework.  Examples of data sets which have been interfaced to BlueOx include the L3 New
Particle Ntuple, the CMS HCAL testbeam data format (in ROOT), and the high energy electron-positron 
collidor format used for demonstrations.

In the discovery phase, the agent determines the available data sets using one more more implementations 
of the Discovery interface.  An implementation can make this determination in many ways, from reading a
local file to querying a central remote database or directly contacting servers to request lists of 
data sets.  This flexibility allows an experiment to use a previously-developed file/run database or to use
a standard implementation such as the LDAP-based discovery classes provided with BlueOx.  The user can 
either request all the data sets, or can request those which match a query on data set name or parameters
specified in the data set descriptor.

The brokering phase is an extremely important phase in the execution of a job.  Brokering assigns a server
for each data set which the user requests for the analysis.  Brokering is handled by the agent through the Brokerage interface.  The brokerage interface accepts a list of data sets and returns a list of contracts
specifying the servers to be used and the data sets to handled by each.  The contracts also specify a 
validity time period for the contract and a 
level of service, including an amount of CPU horsepower to be made available and an estimated delay before
job execution begins.  The contracts are crypographically signed by the issuing authority and the
servers are configured to accept only contracts which can be cryptographically verified.

The communication between the agent and the server during job execution is handled through a 
CommunicationScheme.  Each scheme implements a pair of interfaces, one of which defines access to the
server for the client and the other which defines access to the client for the server.  From the client
side, the interface allows the submission of the job, monitoring of the execution, reception of messages
and exceptions, and access to the completed job fragment.  From the server side, the interface 
allows the server to check the authentication of the user, report messages and exceptions, and return
the completed job.  Servers can offer multiple communication schemes and the clients can select the
scheme to use.  For example, a client behind a firewall might have to use a polling protocol, while a 
directly-connected client might use a protocol allowing a server to directly contact the client.

\subsection{User utilities}

One important interface which is not used for communication or data management is the Mergeable interface.
The agent splits jobs for execution using the standard serialization operations of Java.  The Mergeable
interface supports the reversal of the process, adding together objects to create a merged or summed 
object.  In general, the merge operation is quite easy to define.  For histograms, the merge simply adds 
each histogram to its matching histogram from the other split objects, bin by bin.  For lists, merging
requires simply concatenation (and perhaps sorting) of the lists from the partial analyses.  For compound
objects, each member variable must be merged with its analogue in the other fragments.  

Every analysis must use mergable objects if it wishes to return results from the analysis. To ease the
complexity for the analysis writer, BlueOx provides both a group of mergeable utility classes and an 
automerging facility.  If a class contains only mergeable, static, or transient member variables, it
can be merged automatically using the Java reflection capabilities.  Any object tagged with the 
Automergeable interface is considered to fulfill these requirements and no further work is required
by the author of the class to support the merge operation.

The mergeable classes which BlueOx provides include both container classes (trees and sets) and 
physics-oriented classes including histograms and ntuples. For sets the union operation is used, while in 
the maps values with identical keys are merged themselves.  

Physics analyses depend heavily on histograms and distributions.  BlueOx does not provide these objects
directly, but rather it provides a interface to the standard AIDA (Abstract Interfaces for Data Analysis)
histograms, functions, and ntuples\cite{aida}.  This interface allows any implementation of AIDA (such
as JAIDA\cite{jaida}) to be used with BlueOx.  The transport of the AIDA tree is handled through XML, so
a different AIDA implementation can be used on the server and client without difficulty.

\section{Status}

\begin{figure*}[htb]
\includegraphics[width=\linewidth]{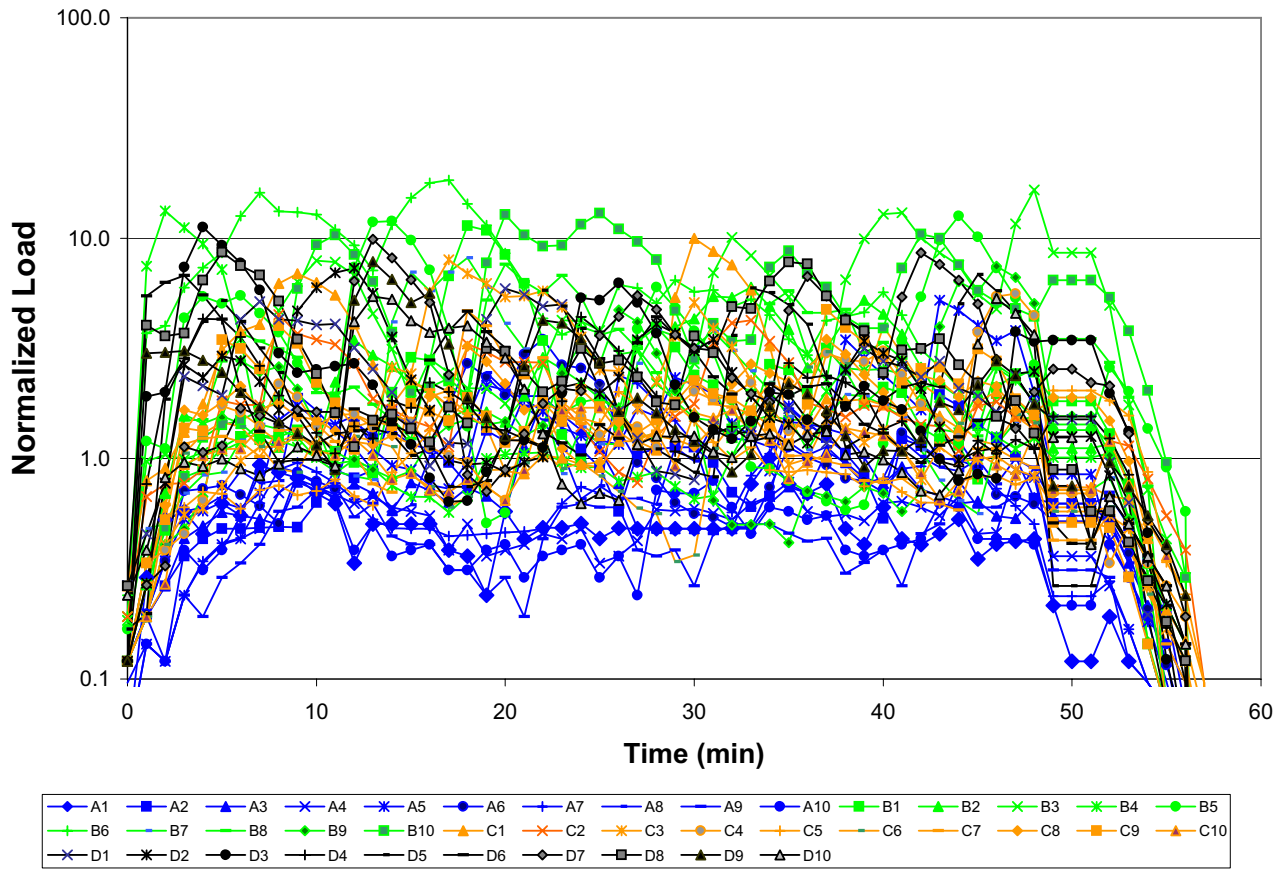}
\caption{Test of the BlueOx system using forty server processes running on four physical machines and
one hundred client processes each sequentially submitting ten jobs to the BlueOx system.  The
dummy data source was used for the test analysis, which simply counted the events which were seen.\label{fig:test}}
\end{figure*}

The BlueOx framework, which was first described in prototype form at CHEP01\cite{chep01}, has been
rewritten to use abstract interfaces and more thoroughly tested than was done for the previous version.
A ``dummy'' data source was written to allow testing of reasonably large-scale BlueOx systems in terms
of server and client process counts without requiring hundreds of physical machines.  For example,
Figure~\ref{fig:test} shows a test with forty server processes running on four physical machines and
one hundred client processes each submitting ten jobs with a random delay between job submissions.  The
brokering process keeps the load on each server reasonably near to one.  In this test, 
the jobs execute much faster (i.e. in milliseconds) than a real analysis job would, which reduces 
somewhat the ability of the servers to load-balance correctly.

For testing and demonstration purposes, we have also written a data source to access linear-collider-like
data generated by Pythia\cite{pythia} then smeared and reconstructed into physics objects by a custom
program.  This data source allows the study and demonstration of BlueOx without involving the release of
any experiment's data.  Using the BlueOx framework, we developed a simple analysis to isolate the
\(\mathrm{e}^+ \mathrm{e}^- \rightarrow \mathrm{ZH} \rightarrow \nu\bar{\nu}\mathrm{b}\bar{\mathrm{b}}\) 
process from background using several analysis techniques including neural networks.

\section{Future Developments}

The BlueOx system is an active object of research and development, as we attempt to improve both the 
operation of the system and our understanding of distributed analysis.  

In the near term, one important center of future study is the brokering system.  In the current default implementation, 
the brokerage uses
seed information stored in an LDAP server about server performance to determine which servers to contact
directly to request contracts.  The agent starts with the server which the seed data indicates would 
provide the highest horsepower for the largest data set, and it requests a contract from that server.  If the
server's offer indicates an available horsepower below that of the next server on the list, the agent will
contact that second server, otherwise it will keep the contract and proceed to the next data set.  The
horsepower offer in the contract is kept and used as the new estimate of the server's availablity.

This process suffers from several problems, including an ``optimistic agent'' problem.  Since the agent 
uses only data about the server's maximum potential horsepower and
not recent load data or user-dependent privileges, the agent may attempt to contact every single server
in the system in search of an improved deal which may not exist.  One solution to the problem is
to provide the agent with more information about the servers' recent load and the amount of horsepower
which would be available to a given user.  Another approach to control contract-related traffic is for a
single computer to be responsible for brokering contacts for a cluster of servers, for example in a farm.

In the longer term, our study targets include the data movement
which will be required to support an active analysis community.
In an active experiment, new data and Monte Carlo are constantly produced, and the focus of analysis
may shift through time.  For BlueOx to be most effective, the analysis should be performed using a local 
cache of the data, if possible.  In an dynamic experiment, the distribution of data among the servers will
have to shift over time as new data and Monte Carlo arrive and analysis groups change interests and
priorities.  While BlueOx is not a data-management system, it can function as an important source of 
information about what data users are requesting and how well the distribution of data matches the
distribution of requests.  

This work was supported by NSF Grant PHY-0219095.


\end{document}